\documentclass[12pt,a4paper]{article}
\usepackage[dvips]{graphicx}      
\usepackage{epsfig}
\psfigdriver{dvips}

\newcommand\be{\begin{equation}}
\newcommand\ee{\end{equation}}
\newcommand\bea{\begin{eqnarray}}
\newcommand\eea{\end{eqnarray}}
\newcommand\ket[1]{|#1\rangle}
\newcommand\bra[1]{\langle #1|}
\newcommand\braket[2]{\langle #1|#2\rangle}
\newcommand{\fatalpha}{{\bf \alpha \kern -0.44em \alpha}}
\newcommand{\fatsigma}{{\bf \sigma \kern -0.54em \sigma}}
\newcommand{\tpchi}{{\bf \chi \kern -0.35em \chi}}
\newcommand{\llambda}{{\bf \lambda \kern -0.45em \lambda}}
\newcommand\tr{\mbox{ Tr }}
\newcommand\ceq[2]{
 \begin{minipage}{#1}{
  \vspace{4pt}
   \begin{center}#2
   \end{center}
  \vspace{-10pt}}
 \end{minipage} }

\newcommand\clebsch[6]{\left<\begin{array}{cc|c}
   #1 & #2 & #3 \\ #4 & #5 & #6 \end{array} \right>}

\newcommand\bsigma{\sigma\hspace{-.56em}\sigma}

\begin{document}
\title{Discrete Moyal-type Representations for a Spin}
\author{Stephan Heiss and Stefan Weigert \\
Institut de Physique, Universit\'e de Neuch\^atel\\ Rue A.-L.
Breguet 1, CH-2000 Neuch\^atel, Switzerland\\ \tt
stefan.weigert@iph.unine.ch}
\date{April 2000}
\maketitle

\begin{abstract}
In Moyal's formulation of quantum mechanics, a quantum spin $s$ is
described in terms of {\em continuous symbols}, i.e. by smooth
functions on a two-dimensional sphere. Such prescriptions to
associate operators with Wigner functions, $P$- or $Q$-symbols,
are conveniently expressed in terms of operator kernels satisfying
the Stratonovich-Weyl postulates. In analogy to this approach, a
{\em discrete} Moyal formalism is defined on the basis of a
modified set of postulates. It is shown that appropriately
modified postulates single out a well-defined set of kernels which
give rise to {\em discrete symbols}. Now operators are represented
by functions taking values on $(2s+1)^2$ points of the sphere. The
discrete symbols contain no redundant information, contrary to the
continuous ones. The properties of the resulting discrete Moyal
formalism for a quantum spin are  worked out in detail and
compared to the continuous formalism, and it is illustrated by the
example of a spin $1/2$.
\end{abstract}

\section{Introduction}

The idea to represent quantum mechanics of a particle in phase
space $\Gamma$ goes back to Wigner \cite{wigner32}. He established
a one-to-one correspondence between a quantum state $\ket{\psi} $
in the particle Hilbert-space ${\cal H}$ and a real function
\be
W_{\psi}(p,q) = \frac{2}{h}\int_\Gamma \, dx \, \psi^* (q+x) \psi
                                                 (q-x) \exp[2ipx/\hbar] \, .
\label{oldwigner} \ee
Its properties suggest an interpretation as a {\em
quasi}-probability in phase space, the only `drawback' being due
to the negative values it may take. A more general framework for
phase-space representations \cite{moyal49} of quantum states as
well as operators ${\widehat A}$  is given by the relation
\be
W_{\widehat A} (q,p) = \mbox{ Tr } \left[ \widehat{ \Delta}(q,p) \widehat
                                A \right] \, ,
\label{newwigner} \ee
with an operator kernel \cite{royer77,huguenin+81}
\be
 \widehat{ \Delta}(q,p) = 2 \widehat{ D} (q,p) \,  {\widehat \Pi } \,
                    \widehat{ D}^\dagger (q,p) \, ,
                    \qquad (q,p) \in \Gamma \, .
\label{particlekernel} \ee
Here ${\widehat \Pi } : (\hat q, \hat p) \to (-\hat q, -\hat p)$
is the unitary, involutive parity operator while $\widehat{ D}
(q,p)$ describes translations in phase space \cite{perelomov86}.
If the operator $\widehat A$ is chosen as the density matrix of a pure
state, $\widehat A = {\hat \rho}_\psi = \ket{\psi} \bra{\psi}$, Eq.\
(\ref{newwigner}) reduces to (\ref{oldwigner}). The kernel $\widehat{
\Delta}(q,p)$ can be derived from a set of conditions which a
phase-space representation is required to satisfy (cf. below). It
is intimately related to the behaviour of a function $W_A
(q,p)$ under {\em translations} mapping the phase space $\Gamma$
onto itself. The map (\ref{newwigner}) from operators to functions
($\widehat A \to W_A$) has an important feature: its inverse, mapping
functions to operators ($ W_A \to \widehat A$), is mediated by the
{\em same} kernel ---in other words, the kernel $\widehat{\Delta}(q,p)$
is {\em self-dual}.

For a quantum spin, the symbol associated with an operator is a
continuous function defined on a sphere ${\cal S}^2$, which is the
phase space of classical spin. Now, instead of translations in
planar phase space, it is the group $SU(2)$ of rotations  which
plays a dominant role when the Moyal formalism is set up. As for a
particle, the set of Stratonovich-Weyl postulates
\cite{stratonovich57}  characterizes the symbols in an elegant
way. For clarity, the postulates are now displayed in their
familiar form for the continuous symbols:
$$
 \begin{array}{lll}
 {(S0)} & {\sf linearity}:&
   \widehat{A} \mapsto W_A ~~~ {\rm is~ a~ linear~ one~ to~ one~ map} , \\
 {(S1)} & {\sf reality}:&
     \begin{minipage}{7cm}  $ W_{A^\dagger}({\bf n}) = W_A^*({\bf n}) \, ,
      $ \end{minipage}  \\
 {(S2)} & {\sf standardisation}:&
     \begin{minipage}{1cm} \vspace{0mm}
        $$ {\strut \frac{2s+1}{4\pi}\int_{{\cal S}^2} W_A({\bf n}) d{\bf n}
         = \tr [ \widehat{A} ] \hfill} \, , $$
     \end{minipage}\\
 {(S3)} & {\sf traciality}: &
     \begin{minipage}{1cm}
           $$ \frac{2s+1}{4\pi}\int_{{\cal S}^2} W_A({\bf n}) W_B({\bf n}) d{\bf n}
             = \tr [ \widehat{A}\widehat{B} ]  \, , \hfill  $$ \vspace{0mm}
     \end{minipage} \\
 {(S4)} & {\sf covariance}: &
     \begin{minipage}{7cm}
        $ W_{g\cdot A} = W_A^g \, , \quad  g \in SU(2)\, . $
     \end{minipage}
 \end{array}
$$
 It is natural to have a {\sf linear} relation between operators
and symbols (S0), while  (S1) implies that hermitean operators are
represented by {\sf real} functions. The third condition (S2) maps
the identity operator to the constant function  on phase space,
and {\sf traciality}  (S3) ensures that the correspondence between
operators and symbols is invertible. The {\sf covariant}
transformation of the symbols with respect to rotations $g \in
SU(2)$ effectively introduces phase-space points as arguments of
the symbols. The continuous Moyal representation for a spin
\cite{varilly+89,amiet+91} compatible with these conditions can be
based on a self-dual kernel $\widehat \Delta({\bf n})$ (cf. Sect.~ 2) in
analogy to (\ref{newwigner}).

In order to have a consistent and full-fledged classical formalism
it is necessary to introduce a product between symbols which keeps
track of the non-commutativity of the underlying operators. This
Moyal product \cite{moyal49}, or twisted product, for two
operators  $ \widehat{A} $ and  $ \widehat{B} $ expresses the
$W_{AB}$ of the operator product $ \widehat{A} \widehat{B} $ in
terms the symbols $W_A$ and $W_B$,
\be \label{eqPMC}
W_{AB} ({\bf n})
     = W_A ({\bf n}) \ast W_B ({\bf n})
     = \int_{{\cal S}^2} \int_{{\cal S}^2}
      L ({\bf n}, {\bf m}, {\bf k}) W_A({\bf m}) W_B({\bf k}) d{\bf m} d{\bf k} \, ,
\ee
with a function $L ({\bf n}, {\bf m}, {\bf k})$ of three arguments
given explicitly in \cite{varilly+89}, for example. The $\ast$
product is known to be associative.

Other continuous representations for a spin do exist, such as the
Berezin symbols of spin operators \cite{berezin75} which are the
analog of the $P$- and $Q$-symbols \cite{perelomov86} for a particle.
Instead of a single self-dual kernel, the Berezin symbols require
however, a {\em pair} of two different kernels, {\em dual} to each
other: one of the kernels maps operators to functions while its
dual is needed for the inverse procedure. It will become clear
later on that the self-dual and the dual approach correspond to
defining orthogonal and non-orthogonal bases, respectively, in the
vector space ${\cal A}_s$ of operators acting on the Hilbert space
of a spin $s$. When slightly modifying the postulates of
Stratonovich, they are also compatible with kernels which are not
self-dual.

A common feature of these representations is the {\em redundancy}
of the continuous symbols. When represented by a $(2s+1)\times
(2s+1)$ matrix, a hermitean operator is fixed by the values of $
(2s+1)^2 $ real parameters. Consequently, the values of the
symbols, continuous functions on the sphere, cannot all be
independent---in other words, the information contained in a
symbol is redundant. The {\em discrete} version of  $ P $-
and  $ Q $-symbols for a spin  $ s $, introduced in
\cite{amiet+98, weigert98/1} as a means to reconstruct the quantum
state of a spin, allows one to characterize a spin operator
$\widehat A$ by using only the {\em minimal} number of parameters.
In fact, a discrete symbol can be considered as living on a
`discretized sphere,' that is, as a function taking (real) values
on a finite set of points on the sphere only. Such a formalism
will be called a discrete Moyal-type formalism.

The purpose of the present paper is to develop the discrete Moyal
formalism in analogy to the continuous one. In particular, the
kernel and its dual defining the discrete symbols will be derived
from a set of appropriate Stratonovich-type postulates.
Subsequently, the properties of these symbols are studied in
detail.

\section{Continuous representations}

\subsubsection*{Continuous self-dual kernel: Wigner symbols}

The Stratonovich-Weyl correspondence for a spin $s$ is a rule
associating with each operator $ \widehat{A} \in {\cal A}_s$ on a
Hilbert space $ {\cal H}_s$ a function  $ W_A $ on the sphere $
{\cal S}^2 $, called its (Wigner-) symbol. Let us define it in
analogy to Eq. (\ref{newwigner}), by means of a universal operator
kernel $\widehat \Delta ({\bf n})$, which can also be thought of
as a field of operators on the sphere. Then, the first requirement
(S0) is already satisfied, and the postulates (S1) to (S4) turn
into conditions on the kernel:
$$
 \begin{array}{lll}
  {(C1)} & {\sf reality}:&
     \begin{minipage}{7cm}
       $ {\widehat \Delta}^\dagger  ({\bf n})  = {\widehat \Delta} ({\bf n})\, ,  $
     \end{minipage}  \\
 {(C2)} & {\sf standardisation}: &
     \begin{minipage}{1cm} \vspace{0mm}
        $$ {\strut \frac{2s+1}{4\pi}\int_{{\cal S}^2}  d{\bf n} \, {\widehat \Delta}
         ({\bf n})           = \hat I \hfill} \, , $$
     \end{minipage}\\
 {(C3)} & {\sf traciality}: &
     \begin{minipage}{1cm}
           $$ \frac{2s+1}{4\pi}\int_{{\cal S}^2} d{\bf n}
          \tr \left[{\widehat\Delta} ({\bf n}) {\widehat \Delta} ({\bf m})\right]
                                  {\widehat \Delta} ({\bf n})
                  = {\widehat \Delta} ({\bf m}) \, , \hfill  $$ \vspace{0mm}
     \end{minipage} \\
 {(C4)} & {\sf covariance}: &
     \begin{minipage}{7cm}
     $ {\widehat \Delta} (g \cdot {\bf n})
          = {\widehat U}_g \, {\widehat \Delta} ({\bf n})\,  {\widehat U}_g^\dagger
          \, , \quad  g \in SU(2)\, . $
    \end{minipage}
 \end{array}
$$
where the matrices ${\widehat U}_g $ are a unitary
$(2s+1)$-dimensional irreducible representations of the group
$SU(2)$.

The existence of a kernel ${\widehat \Delta ({\bf n}) }$
satisfying (C1-4) has been proven in \cite{stratonovich57} by
explicit construction. The derivation in \cite{varilly+89} starts
by expanding the kernel in a basis associated with the eigenstates
of the operator ${\bf \hat{s}}\cdot{\bf n}_z$,
\be \label{varillyansatz}
{\widehat \Delta ({\bf n}) }
     = \sum_{m,m'=-s}^s Z_{mm'} ({\bf n}) \ket{m,{\bf n}_z} \bra{m', {\bf n}_z} \, ,
\ee
with unknown coefficients $Z_{mm'} ({\bf n})$. It follows from
(C1-4) that one must have
\be \label{eqSolBrute}
  Z_{mm^\prime}^s ({\bf n})
         = \frac{\sqrt{4\pi}}{2s+1} \sum_{l=0}^{2s} \varepsilon_l \sqrt{2l+1}
             \clebsch{s}{l}{s}{m}{m^\prime-m}{m^\prime}
              Y_{l, m^\prime-m}({\bf n}) \, ,
\ee
where  $ \varepsilon_0=1 $ and $ \varepsilon_l=\pm 1 \, , l=1,
\ldots, 2s$, and the definition of Clebsch-Gordan coefficient given in \cite{varilly+89} is used. Consequently, there are $ 2^{2s} $ different kernels
which define a Stratonovich-Weyl correspondence rule.

A new and simple derivation of the kernel ${\widehat \Delta ({\bf
n}) } $, independent of the argument given in \cite{varilly+89},
is presented now which has two important advantages. On the one hand,
it will provide a form of the kernel similar to that one of a
particle (\ref{particlekernel}), which is interesting from a
conceptual point of view. On the other hand, it will be possible
to transfer this approach to a large extent to the case of the
discrete Moyal formalism.

Expand the kernel in the eigenbasis of the operator $ \hat{\bf
s}\cdot {\bf n} $,
\be \label{eqDevNoy}
  \widehat{\Delta}({\bf n})
        = \sum_{m, m^\prime=-s}^s \Delta_{mm^\prime} ({\bf n})
              \ket{m, {\bf n}}   \bra{m^\prime, {\bf n}} \, ,
\ee
where the expansion coefficients $\Delta_{mm^\prime}$ are unknown so far.
According to the reality condition (C1) they must satisfy $\Delta_{mm^\prime}
({\bf n}) = \Delta_{m^\prime m}^* ({\bf n})$. In a first step, the
numbers $ \Delta_{mm^\prime} ({\bf n}) $ are shown {\em not} to
depend on the label ${\bf n}$. Consider the transformation of
$\widehat{\Delta}({\bf n})$ under a rotation $g$. According to
(C4) one must have
\bea
\nonumber \sum_{m, m^\prime=-s}^s
& & \hspace{-2em}
      \Delta_{mm^\prime} (g\cdot{\bf n})
              \ket{m, g \cdot {\bf n}}   \bra{m^\prime, g \cdot {\bf n}} =  \\
&=&  {\widehat U}_g \, \widehat \Delta ({\bf n}) \, {\widehat U}_g^\dagger
= \sum_{m, m^\prime=-s}^s \Delta_{mm^\prime} ({\bf n})
              \ket{m, R_g{\bf n}}   \bra{m^\prime, R_g {\bf n}} \, ,
\label{firststep}
\eea
where ${\widehat U}_g \ket{m, {\bf n}} = \ket{m, R_g{\bf n}} =
\ket{m, g\cdot {\bf n}}$ with a rotation matrix $R_g \in SO(3)$
representing $g \in SU(2)$ in $I \!\! R^3$. Consequently, one must
have
\be
\Delta_{mm^\prime} (g\cdot {\bf n})  = \Delta_{mm^\prime} ({\bf n}) \, ,
%
\label{invar1delta}
\ee
which is only possible if $\Delta_{mm^\prime}$ does not depend on
${\bf n}$. Consider next a rotation $ g({\bf n}) $ about the axis
${\bf n}$ by an angle $\varphi \in [0,2\pi)$, represented by the
unitary $ {\widehat U}_{g({\bf n})}  = \exp(i {\bf n} \cdot
\hat{\bf s} \varphi) $. The left-hand-side of (C4) is  invariant
under this transformation while the right-hand-side transforms:
\be
 \widehat{\Delta}(R_{g ({\bf n})} {\bf n})
     = \widehat{\Delta}( {\bf n})
     =  \sum_{m, m^\prime=-s}^s \Delta_{mm^\prime} \exp[i(m-m') \varphi]
              \ket{m, {\bf n}}   \bra{m^\prime, {\bf n}}  \, ,
\label{invar2delta}
\ee
which is possible only if
\be
\Delta_{m m^\prime} = \Delta(m) \delta_{m m^\prime} \, .
\label{diagonaldelta}
\ee
Therefore, covariance of the kernel under elements of $SU(2)$
requires it to be diagonal in the basis associated with the
direction ${\bf n}$,
\be
\widehat{\Delta} ({\bf n})
   = \sum_{m=-s}^s \Delta (m) \ket{m, {\bf n}} \bra{m, {\bf n}} \, .
\label{eqNoyDiscECG}
\ee
Next, the condition of traciality will be exploited. Upon
rewriting (C3) in the form
\be
\int_{{\cal S}^2} d{\bf n} \, \delta_s ({\bf m}, {\bf n}) \widehat
\Delta ({\bf n}) = \widehat \Delta ({\bf m}) \, ,
\label{C3'}
\ee
the function $ \delta_s ({\bf m}, {\bf n}) \equiv (2s+1) \tr
[\widehat\Delta({\bf m}) \widehat\Delta({\bf n})] /(4\pi) $ is seen to be
the {\em reproducing kernel} for a certain subset of $(2s+1)^2$ functions
on the sphere \cite{arecchi+72,varilly+89}. In other words, $ \delta_s({\bf m},
{\bf n})$ acts in this space as a delta-function with respect to integration
over ${\cal S}^2$, and for spin $s$, it reads explicitly
\be
  \delta_s ({\bf m}, {\bf n})
  = \sum_{l=0}^{2s} \sum_{m=-l}^l Y_{lm}({\bf m}) Y^*_{lm}({\bf n})
  = \sum_{l=0}^{2s} \frac{2l+1}{4\pi} P_l({\bf m}\cdot {\bf n}) \,
  .
\label{repro}
\ee
Here the addition theorem for spherical harmonics
 $Y_{lm}({\bf n}), l=0, \ldots, 2s, -l\leq m \leq l$,
 has been used to express the sum over $m$ in terms of Legendre polynomials
$P_l(x), -1\leq x \leq 1$. Upon choosing $ {\bf m} \equiv {\bf n}_z $
and with $ {\bf n}_z\cdot{\bf n} = \cos\theta$, the condition (C3) becomes
\be
\tr \left[ \widehat\Delta({\bf n}_z) \widehat\Delta({\bf
n})\right]
  = \sum_{l=0}^{2s} \frac{2l+1}{2s+1} P_l(\cos\theta)\, .
\label{tracial1}
\ee
Use now the expansion (\ref{eqNoyDiscECG}) of the kernel on the
left-hand-side as well as the identity 
\be
\nonumber |\braket{m, {\bf n}_z}{m^\prime, {\bf n}}|^2 =
      \sum_{l=0}^{2s} \frac{2l+1}{2s+1} \clebsch{s}{l}{s}{m}{0}{m}
              \clebsch{s}{l}{s}{m^\prime}{0}{m^\prime} P_l(\cos\theta) \, ,
\label{eqNormeCoherentGen}
\ee
leading to
 \be
\nonumber \tr \left[ \widehat\Delta({\bf n}_z)
\widehat\Delta({\bf n}) \right]
     = \sum_{l=0}^{2s}
        \left(\sum_{m=-s}^s \Delta (m) \clebsch{s}{l}{s}{m}{0}{m} \right)^2                                  \frac{2l+1}{2s+1}P_l  (\cos\theta) \, .
\label{tracial2}
\ee
Compare now the coefficients of the Legendre polynomials $P_l
(\cos\theta)$ with those in Eq.\ (\ref{tracial1}). This leads to
$(2s+1)$ conditions
\be
\sum_{m=-s}^s \Delta (m) \clebsch{s}{l}{s}{m}{0}{m}
      = \varepsilon_l \, , \quad \varepsilon_l = \pm 1 \, , \quad l = 0, \ldots, 2s \, .
\label{eqBaseF}
\ee
These equations can be solved for $\Delta (m)$ by means of an
orthogonality relation for Clebsch-Gordan coefficients
\cite{edmonds57},
\be
 \Delta (m)
      = \sum_{l=0}^{2s} \varepsilon_l \frac{2l+1}{2s+1}
      \clebsch{s}{l}{s}{m}{0}{m} \, .
\label{eqSolfm1}
\ee
Thus, the self-dual kernel for the continuous Moyal formalism is given by
\be
\widehat\Delta({\bf n})
         = \sum_{m=-s}^s \sum_{l=0}^{2s} \varepsilon_l \frac{2l+1}{2s+1}
                 \clebsch{s}{l}{s}{m}{0}{m} \ket{m, {\bf n}}\bra{m, {\bf n}} \, ,
\label{eqCinquieme}
\ee
Out of these $ 2^{2s+1} $ distinct solutions only $2^{2s}$ are
compatible with the condition of standardization (C2) which has
not been used until now. This condition imposes
\be
\sum_{m=-s}^s \Delta (m) =1 \, ,
\label{standard}
\ee
being satisfied if and only if $ \varepsilon_0 = +1 $.

The set of solutions (\ref{eqCinquieme}) coincides indeed with
those found in \cite{varilly+89}. The easiest way to see this is
to calculate the matrix elements of the kernel $\widehat
\Delta({\bf n}) $ in (\ref{eqCinquieme}) with respect to the
standard basis $\ket{m, {\bf n}_z}$. One reproduces the
coefficients of the expansion (\ref{eqSolBrute}):
$ \bra{m, {\bf n}_z} \widehat \Delta ({\bf n}) \ket{m^\prime, {\bf n}_z}
         = Z_{mm^\prime}({\bf n})$.

The expansion (\ref{eqCinquieme}) is interesting from a conceptual
point of view. It allows one to interpret physically the kernel
$\widehat\Delta({\bf n})$ in analogy with the kernel for a
particle given in (\ref{particlekernel}) by writing
\be \label{suchaniceform}
\widehat\Delta({\bf n})
            = \widehat U_{\bf n}  \widehat\Delta({\bf n}_z)
                           \widehat U_{\bf n}^\dagger \, ,
\label{beforecontraction}
\ee
where $\widehat U_{\bf n}$ represents a rotation which maps the
vector ${\bf n}_z$ on ${\bf n}$. Imagine now to contract
\cite{arecchi+72} the group $SU(2)$ to the Heisenberg-Weyl group.
It is known this procedure turns rotations  $\widehat U_g$ into
translations $\widehat D (q,p)$ . As shown in \cite{amiet+00/1},
the operator $\widehat\Delta({\bf n}_z)$ contracts in the
following way,
\be
\widehat\Delta({\bf n}_z) \to 2 \widehat \Pi \, ,
\label{contract!}
\ee
if $\varepsilon_l = +1, l= 0,\ldots, 2s$. Therefore, the operator
$\widehat\Delta({\bf n}_z) $ plays the role of parity for a spin
which is by no means immediately obvious when looking at it.

Finally, we would like to point out that the integral kernel $L$,
defining the $\ast$ product of symbols (\ref{eqPMC}), has a simple expression in
terms of Wigner kernels:
\be \label{eqNoyauPMC}
L ({\bf n}, {\bf m}, {\bf k})
       = \left( \frac{2s+1}{4\pi} \right)^2
        \tr \left[{\widehat \Delta}({\bf n})
        {\widehat \Delta}({\bf m}) {\widehat \Delta}({\bf k}) \right] \, .
\ee
\subsection*{Continuous dual kernels: Berezin symbols}
Wigner symbols of spin operators are calculated by means of a
kernel which is its own dual. Other phase-space representations
are known which do not exhibit this `symmetry' between an operator
and its symbol. $P$- and $Q$-symbols for a particle are familiar
examples which have their analog in the `Berezin' symbols for a
spin. It will be shown now that these symbolic representations
also have a simple description in terms of kernels satisfying a
modified set of Stratonovich-Weyl postulates. The conditions
(C1-4) must be relaxed slightly in order to allow for a pair of
dual kernels.

The required generalization is easily understood in terms of
linear algebra. The ensemble of all operators, that is, the
self-dual kernel is nothing but a an (overcomplete) set of vectors
spanning the linear space ${\cal A}_s$ of operators on the Hilbert
space of the spin $s$. As the traciality (C3) indicates, this
family of vectors is `orthogonal' with respect to integration over
the sphere as a scalar product. Each operator $\widehat A$ can be
written as a linear combination of the elements of the kernel with
its Wigner symbol as expansion coefficients. More precisely, the
expansion coefficients $W_A ({\bf n})$ with respect to the basis
$\widehat \Delta ({\bf n}) $ are given by the `scalar product' of
$\widehat A$ with the {\em same} basis vector as shown, for
example, in Eq.\ (\ref{newwigner}). The essential point now is,
that there are also {\em non-orthogonal} bases of the same space.
Given a non-orthogonal basis, denoted by ${\widehat \Delta}_{\bf
n}$, its {\em dual} basis ${\widehat \Delta}^{\bf n}$ is uniquely
determined through the scalar product. Furthermore, the dual basis
also spans the original space which implies that now there will be
{\em two} different expansions of one operator $\widehat A$
defining a symbol $A_{\bf n}$ and its dual $A^{\bf n} $.
Consequently, both kernels and symbols now come in pairs. The
familiar $P$- and $Q$-symbols---or Berezin symbols
\cite{berezin75}---will turn out to be related in this precisely
way.

Non-orthogonal bases are allowed in the present framework if,
first of all, traciality (C3) is relaxed to
 \be
{(C'3)} \quad {\sf traciality}: \qquad
      \frac{2s+1}{4\pi}\int_{{\cal S}^2} d{\bf n}
      \tr \left[ {\widehat\Delta}_{\bf m} {\widehat \Delta}^{\bf
          n}\right]  {\widehat \Delta}_{\bf n}
      = {\widehat \Delta}_{\bf m} \, .
\ee
The kernel and its dual are both real in analogy to (C1).
Explicitly, the symbols and their duals are given by
\be \label{dualsyms}
A_{\bf n} = \tr \left[ \widehat A {\widehat \Delta}_{\bf n}
\right] \, ,
          \qquad
A^{\bf n} = \tr \left[ \widehat A {\widehat \Delta}^{\bf n}
\right] \, .
\ee
Furthermore, one is free to normalize one of the kernels,
${\widehat \Delta}_{\bf n}$, say, in analogy to (C2), and one
requires it to transform covariantly (C4).

It is possible as before to derive the explicit form of the
kernels by a reasoning in analogy to above. The general ansatz for
both ${\widehat \Delta}_{\bf n}$ and $ {\widehat \Delta}^{\bf n}$
in the basis referring to the axis ${\bf n}$ as in
(\ref{eqDevNoy}) is again reduced to diagonal form by exploiting
their behaviour under rotations:
\be
\widehat{\Delta}_{{\bf n}}
 = \sum_{m=-s}^s \Delta_m \ket{m,{\bf n}}\bra{m,{\bf n}} \, ,
 \qquad
 \widehat{\Delta}^{{\bf n}}
 = \sum_{m=-s}^s \Delta^m \ket{m,{\bf n}}\bra{m,{\bf n}}\, ,
\label{dualdiag} \ee
with two sets of numbers $\Delta_m$ and $\Delta^m$, which do not
depend on ${\bf n}$. It is necessary that the trace of these two
operators with labels ${\bf m} \equiv {\bf n}_z$ and ${\bf n}$,
say, equals the reproducing kernel with respect to integration
over the sphere, that is, instead of (\ref{tracial1}) one needs to
have
\be
\tr \left[ {\widehat\Delta}_{{\bf n}_z} {\widehat \Delta}^{\bf n}
\right]
  = \sum_{l=0}^{2s} \frac{2l+1}{2s+1} P_l(\cos\theta)\, ,
\label{tracial3} \ee
where $\cos\theta \equiv {\bf n}_z \cdot {\bf n}$. This leads to
the conditions
\be
       \left[ \sum_{m=-s}^s \Delta_m \clebsch{s}{l}{s}{m}{0}{m} \right]
       \left[ \sum_{m=-s}^s \Delta^m \clebsch{s}{l}{s}{m}{0}{m} \right] = 1 \, ,
\label{dualcond}\ee
with $l= 0, \ldots, 2s$. The ensemble of solutions is
parameterized by $(2s+1)$ non-zero real numbers $\gamma_l$:
\be
      \sum_{m=-s}^s \Delta_m \clebsch{s}{l}{s}{m}{0}{m} =
      \left[ \sum_{m=-s}^s \Delta^m \clebsch{s}{l}{s}{m}{0}{m}\right]^{-1} =
      \gamma_l \, ,
\label{constraintnew} \ee
Solving for the expansion coefficients, one obtains
 \bea
      \Delta_m &=& \sum_{l=0}^{2s} \gamma_l \frac{2l+1}{2s+1}
              \clebsch{s}{l}{s}{m}{0}{m} \, , \\
      \Delta^m &=& \sum_{l=0}^{2s} \gamma_l^{-1} \frac{2l+1}{2s+1}
              \clebsch{s}{l}{s}{m}{0}{m} \, .
\label{dualresult} \eea
As in the self-dual case, the standardisation implies that $\gamma_0=+1$.
This class of solutions for kernels which are not their own dual has been
obtained in \cite{amiet+91} by an entirely different approach. Self-dual kernels are a small subset: they require $\Delta_m\equiv \Delta^m$,
which is $\gamma_l = \gamma_l^{-1}$ or $\gamma_l = \pm 1$ in agreement with
(\ref{eqCinquieme}). Each set of numbers $\gamma_l$ defines a consistent
phase-space representation of a quantum mechanical spin.

Consider the particular case $ \Delta_m = \delta_{ms} $ resulting
from
\be
\gamma_l =  \clebsch{s}{l}{s}{s}{0}{s} \, .
\label{choicePQ}\ee
The associated kernels read
\bea \label{contcohpro}
    \widehat{\Delta}_{\bf n} &=& \ket{s, {\bf n}}\bra{s, {\bf n}}
                               \equiv \ket{{\bf n}}\bra{{\bf n}}\, , \\
    \widehat{\Delta}^{\bf n} &=& \sum_{m=-s}^s
     \sum_{l=0}^{2s} \frac{2l+1}{2s+1}\clebsch{s}{l}{s}{s}{0}{s}^{-1}
    \clebsch{s}{l}{s}{m}{0}{m}   \ket{m,{\bf n}}\bra{m,{\bf n}} \, .
        ,
\label{PQkernels} \eea
This choice has the advantage that one of the two symbols,
reducing to the expectation value of an operator $\widehat A$ in
coherent states \cite{perelomov86}, is particularly simple. It turns out to be just
the $Q$-symbol, $Q_A ({\bf n}) = \bra{{\bf n}} \widehat
A \ket{{\bf n}}$, that is, its expectation in a spin-coherent
state. At the same time, one falls back on a familiar expression
for the dual symbol which turns out to be the $P$-symbol for
$\widehat A$, defined by an expansion in terms of a linear
combination of operators projecting on coherent states,
\be
\widehat{A} = \frac{(2s+1)}{4\pi} \int_{{\cal S}^2} P_A({\bf n})
\ket{{\bf n}}\bra{{\bf n}} d{\bf n} \, .
\label{Psymbol} \ee
In the present notation one simply has in view of (\ref{dualsyms})
that
\be \label{PQasduals}
Q_A({\bf n}) \equiv A_{\bf n}
  = \tr [ \widehat A \widehat{\Delta}_{\bf n}]  \, ,
  \qquad
P_A({\bf n}) \equiv A^{\bf n}
  = \tr [ \widehat A \widehat{\Delta}^{\bf n}] \, ,
\ee
so that (\ref{Psymbol}) reads
\be \label{Psymbol2}
\widehat{A}
   = \frac{(2s+1)}{4\pi} \int_{{\cal S}^2} d{\bf n}
       \tr [ \widehat A \widehat{\Delta}^{\bf n}] \,
                 \widehat{\Delta}_{\bf n} \, .
 \ee
It is obvious now that one has (cf. \cite{varilly+89})
\be \label{thisandthat}
\tr [ \widehat A \widehat B ]
   = \frac{(2s+1)}{4\pi} \int_{{\cal S}^2} d{\bf n} \,
        P_A({\bf n}) Q_B({\bf n})
   = \frac{(2s+1)}{4\pi} \int_{{\cal S}^2} d{\bf n} \,
       A^{{\bf n}} B_{{\bf n}}    \, .
\ee
Finally, it is interesting to calculate the $Q$- and $P$-symbols
of the self-dual kernel $\widehat \Delta ({\bf n})$ as well as
the pair of dual kernels $\widehat{\Delta}^{\bf n}$ and
$\widehat{\Delta}_{\bf n}$ using the short-hand
\be \label{short1}
{\sf Y}_{lm}({\bf n}, {\bf m})
     = Y^*_{lm} ({\bf n}) Y_{lm}({\bf m}) \, ,
\ee
so that the reproducing kernel is given by
\be
\delta_{s} ({\bf n}, {\bf m})
 = \sum_{ml} {\sf Y}_{lm}({\bf n}, {\bf m})
  \, .
\label{short2} \ee
\vspace{5mm}
\begin{center}
\begin{tabular}{p{1cm}|p{6.5cm}|p{5.6cm}|}
 & \ceq{6.5cm}{$\tr [ \; \cdot \; \widehat{\Delta}_{\bf m} ] $}
 & \ceq{5.5cm}{$\tr [ \;  \cdot \; \widehat{\Delta}^{\bf m} ]$} \\
 \hline
 \ceq{1cm}{$\widehat{\Delta}_{\bf n} $}
 & \ceq{6.5cm}{$\frac{4\pi}{2s+1}
    \sum_{ml} \clebsch{s}{l}{s}{m}{0}{m}^2 {\sf Y}_{lm}({\bf n}, {\bf m})$}
 & \ceq{5.5cm}{$\delta_{s} ({\bf n}, {\bf m})$} \\
 \hline
 \ceq{1cm}{$\widehat{\Delta}^{\bf n} $}
 & \ceq{6.5cm}{$\delta_{s} ({\bf n}, {\bf m})$}
 & \ceq{5.5cm}{$\sum_{ml}\clebsch{s}{l}{s}{m}{0}{m}^{-2} {\sf Y}_{lm}({\bf n}, {\bf m})$}
  \\
 \hline
   \ceq{1cm}{$\widehat \Delta ({\bf n})$}
 & \ceq{6.5cm}{$\sum_{ml}\clebsch{s}{l}{s}{m}{0}{m}
{\sf Y}_{lm}({\bf n}, {\bf m})$}
 & \ceq{5.5cm}{$\sum_{ml}\clebsch{s}{l}{s}{m}{0}{m}^{-1} {\sf Y}_{lm}({\bf n},
{\bf m})$} \\
 \hline
\end{tabular}
\end{center}
\vspace{5mm}
Note that the entries of last row, the $Q$- and $P$-symbols of the
self-dual kernel $\widehat \Delta ({\bf n})$, do simultaneously
provide the Wigner symbols of the dual kernels
$\widehat{\Delta}_{\bf m} $ and $\widehat{\Delta}^{\bf m} $.
\section{Discrete Moyal-type representations}

A particular feature of the kernels discussed so far is their
redundancy: the linear space of hermitean operators for a spin $s$
has dimension $(2s+1)^2$ while the kernels consist of a
continuously labeled set of basis vectors. In other words, there
are at most $N_s = (2s+1)^2$ linearly independent operators among
all $\widehat \Delta ({\bf n})$, ${\bf n} \in {\cal S}^2$. In this
section {\em discrete} kernels will be introduced, denoted by
$\widehat\Delta_\nu $, $ \nu=1,\ldots,  N_s $. No linear relations
must exist between the operators $\widehat\Delta_\nu $ which
constitute the kernel, that is, they are a basis of ${\cal A}_s$
in the strict sense. It is natural to expect that a {\em subset}
of precisely $N_s$ operators $\widehat \Delta ({\bf n}_\nu),$ $\nu =
1, \ldots , N_s$ will give rise to a discrete kernel. Therefore,
evaluating a continuous symbol of an operator $\widehat A$ at $
N_s $ points $ {\bf n}_\nu $  of the sphere  $ {\cal S}^2 $,
provides a promising candidate for a discrete symbol, i.e. the set $A_\nu
\equiv A_{{\bf n}_\nu}$, $\nu=1, \ldots, N_s $. For brevity, $ N_s $ points on $ {\cal
S}^2 $ are called a {\em constellation}.

As before, one might expect orthogonal and non-orthogonal kernels
to exist. It turns out, however, that an appropriately modified
set of Stratonovich-Weyl postulates covering discrete kernels does
{\em not} allow for orthogonal ones. Therefore we start
immediately by deriving the discrete non-orthogonal kernels coming
as before in combination with a dual.

\subsection*{Discrete dual kernels}

By analogy with the continuous representation of the preceding
section, one modifies the Stratonovich-Weyl postulates in the
following way (throughout the index $\nu$ takes  all the values
from $1$ to $N_s$):
$$
 \begin{array}{lll}
  {(D0)} & {\sf linearity:} &
   \widehat{A} \mapsto A_\nu  \qquad \mbox{ is a linear map} \, , \\
  {(D1)} & {\sf reality:}&
     \begin{minipage}{7cm}
       $ \widehat\Delta^\dagger_\nu = \widehat\Delta_\nu,~~ \nu=1,\ldots,N_s \, ,  $
            \end{minipage}  \\
 {(D2)} & {\sf standardisation:} &
     \begin{minipage}{1cm}  \vspace{0mm}
    $$ \frac{1}{2s+1} \sum_{\nu=1}^{N_s} \widehat\Delta^\nu = \hat{I} \, ,  $$
     \end{minipage}\\
 {(D3)} & {\sf traciality:} &
     \begin{minipage}{1cm}
          $$ \widehat\Delta_\nu = \frac{1}{2s+1} \sum_{\mu=1}^{N_s}
      \tr \left[ \widehat\Delta_\nu \widehat\Delta^\mu\right] \widehat\Delta_\mu  $$
   \vspace{0mm}
     \end{minipage} \\
 {(D4)} & {\sf covariance:} &
     \begin{minipage}{7cm}
     $ {\widehat \Delta}_{g \cdot \nu}
          = {\widehat U}_g \, {\widehat \Delta}_\nu \, {\widehat U}_g^\dagger\, ,
          \quad  g \in SU(2)\, . $
    \end{minipage}
 \end{array}
$$
Let us briefly comment on these conditions. {\sf Linearity} is
automatically satisfied if discrete symbols are defined via
kernels, that is, $ A_\nu  = \tr [ \widehat{A} \widehat\Delta_\nu
] $. The second condition, {\sf reality},  is obvious, and in (D2)
the kernel $\widehat\Delta^\nu$ {\em dual} to $\widehat\Delta_\nu$
is standardized. The duality between $\widehat\Delta_\nu$ and
$\widehat\Delta^\nu$ is made precise by the condition of
traciality since (D3) only holds if one has
\be
   \frac{1}{2s+1}
     \tr \left[ \widehat\Delta_\nu \widehat\Delta^\mu \right]
         = \delta_\nu^\mu,
             \qquad \nu, \mu = 1,\ldots, N_s \, ,
\label{eqDefOrthoBD} \ee
which, upon considering the trace as a scalar product, is
precisely the condition defining the dual of a given basis. As a
matter of fact, if $ \{\widehat{\Delta}_\nu \}$, $\nu = 1, \ldots,
N_s$, {\em is} a basis, its unique dual is guaranteed to exist.
Finally, {\sf covariance} under rotations $g \in SU(2)$ must be
reinterpreted carefully. Under a transformation $g $, a
constellation associated with $N_s$ points on the sphere will, in
general, be mapped to a {\em different} constellation. In other
words, the image $\widehat\Delta_{g\cdot \nu} = \widehat\Delta (g
\cdot {\bf n}_\nu)$ is typically {\em not} one of the operators
$\widehat\Delta_\nu$. Nevertheless, condition (D4) is not empty:
for appropriately chosen rotations $g_{\nu\mu}$ one can indeed map
an operator defined at ${\bf n}_\nu$ to another one associated
with the point ${\bf n}_\mu$, say. In this case, the consequences
for the coefficients of the operators $\widehat\Delta_\nu$ and
$\widehat\Delta_\mu$ are identical to those obtained in the
continuous case. Similarly, invariance of the operator
$\widehat\Delta_\nu$  under a rotations about the axis ${\bf
n}_\nu$ has the same impact as before. Thus the general ansatz for
the discrete kernel (obtained from (\ref{eqDevNoy}) by setting
${\bf n} \to {\bf n}_\nu$) is reduced by exploiting the postulates
(D1-4) to the form
\be
\widehat\Delta_\nu
   \equiv \widehat{\Delta} ({\bf n}_\nu)
   = \sum_{m=-s}^s \Delta_m \ket{m,{\bf n}_\nu} \bra{m,{\bf n}_\nu}
   \, , \qquad \nu = 1, \ldots, N_s \, .
\label{eqNoyDiscECGD} \ee
Therefore, the discrete kernel $\widehat\Delta_\nu$ can be thought
of as a subset of $N_s$ operators $\widehat{\Delta} ({\bf
n}_\nu)$, each one associated with a point ${\bf n}_\nu$ of the
sphere.

Let us mention an important difference between discrete and
continuous kernels, $\widehat\Delta_\nu$ and $\widehat{\Delta}
({\bf n})$, which arises in spite of their formal similarity. Once
the coefficients $\Delta_m$ are fixed a continuous kernel is
determined completely. Discrete kernels, however, come in a much
wider variety since they depend, in addition, on the selected
constellation of points on the sphere. The discrete kernel does
not enjoy the $SU(2)$ symmetry in the same way as does the
continuous one. The discrete subgroups of $SU(2)$ being limited in
type, the continuous symmetry will usually not be turn into a
discrete one. Note, further,  that the elements of the dual kernel
depend, in general, on {\em all} the points of the constellation:
$ \widehat\Delta^\nu = \widehat\Delta^\nu ({\bf n}_1..., {\bf
n}_{N_s})$. This is easily seen from (\ref{eqDefOrthoBD}) since
the variation of a single $ \widehat{\Delta}_\nu $ will have an
effect on all $ \{\widehat{\Delta}^\nu \}$ in order to maintain
orthogonality.

The additional freedom of selecting specific constellations is
connected to a subtle point: actually, not all constellations of
$N_s$ points give rise to a basis in the space ${\cal A}_s$. This
remark is easily understood by considering $I\!\!R^3$ as an
example of a linear space. The (continuous) collection of all unit
vectors in three-space clearly spans it while not every subset of
three vectors is a basis---they might lie in a plane. By analogy,
one must ensure that the operators $\widehat\Delta_\nu, \nu = 1,
\ldots, N_s$, associated with a specific constellation, do indeed
form a basis of ${\cal A}_s$. The operators are indeed linearly
independent if the determinant of their (positive definite and
symmetric) Gram matrix ${\sf G}$ \cite{greub63} satisfies
\be \label{eqDefG}
    \det {\sf G} > 0\, , \qquad {\sf G}_{\nu \nu^\prime}
   = \tr \left[ \widehat\Delta_\nu \widehat \Delta_{\nu^\prime} \right] \, ,
\ee
a condition, which will be studied later in more detail.

Suppose now that the $N_s$ operators $\widehat\Delta_\nu$ in
(\ref{eqNoyDiscECGD}) do form a basis. Then, the kernel dual to
it, that is the set of operators $\widehat\Delta^\nu$, is
determined by the condition (\ref{eqDefOrthoBD}) instead of Eq.\
(\ref{tracial2}). Therefore, one cannot proceed as before to
derive the conditions (\ref{dualcond}). In particular, it is no
longer true that the elements of the dual kernel have an expansion
analogous to (\ref{eqNoyDiscECGD}). This follows immediately
from the impossibility to satisfy (\ref{dualcond}) by an
ansatz for $\widehat\Delta^\nu$ of the form (\ref{eqNoyDiscECGD}):
Eq. (\ref{eqNoyDiscECGD}) represent $N_s$ conditions but a dual of the form
depends only on $(2s+1)$ free parameters $\Delta^m$. Nevertheless, a dual
kernel $\widehat\Delta^\nu$ does exist and it is determined unambiguously---it simply cannot have the form (\ref{eqNoyDiscECGD}). ({\em A also ori}, there is no {\em self-dual} kernel associated with the Stratonovich-Weyl postulates (D0-4)). Consequently, one expands any (self-adjoint) operator $\hat A$ either in terms of a given
kernel,
\be \label{discexp}
         \widehat{A}=\frac{1}{2s+1} \sum_{\nu=1}^{N_s} A^\nu
         \widehat\Delta_\nu\, , \qquad
         A^\nu = \tr \left[ \hat A \widehat\Delta^\nu \right] \, ,
\ee
or, equivalently, in terms of the dual kernel,
\be \label{discdualexp}
         \widehat{A}=\frac{1}{2s+1} \sum_{\nu=1}^{N_s} A_\nu
         \widehat\Delta^\nu\, , \qquad
         A_\nu = \tr \left[ \hat A \widehat\Delta_\nu \right] \, .
\ee
The collection $A \equiv ( A_1, \ldots, A_{N_s})$ of real
coefficients in (\ref{discdualexp}) now is defined as the {\em
discrete phase-space symbol} of the operator $\widehat A$, and
$A^{dual} \equiv ( A^1, \ldots, A^{N_s})$ is the dual symbol.

The relation between the discrete symbol and its dual as well as
between the pair of kernels is linear. It is easily implemented by
means of the Gram matrix ${\sf G}$ and its inverse ${\sf G}^{-1}$,
\be \label{eqDefGinv}
    \left({\sf G}^{-1} \right)_{\nu \nu^\prime}
    \equiv {\sf G}^{\nu \nu^\prime}
   = \frac{1}{(2s+1)^2} \tr \left[ \widehat\Delta^\nu \widehat
     \Delta^{\nu^\prime} \right]
  \, .
\ee The  matrix ${\sf G}$ thus plays the role of a metric,
\be \label{forth}
\widehat{\Delta}^\nu
    = (2s+1) \sum_{\mu=1}^{N_s} {\sf G}^{\mu\nu} \widehat{\Delta}_\mu \, ,
\ee
and the dual symbol is determined according to
\be \label{discrPgen}
   A^\nu =
     (2s+1) \sum_{\nu^\prime=1}^{N_s} {\sf G}^{\nu \nu^\prime} A_{\nu^\prime}
     \, . 
\ee
The trace of two operators $\widehat A$ and $\widehat B$ is easily
found to be expressible as a combination of a discrete symbol and
a dual one,
\be
   \tr \left[\widehat A \widehat B \right]
   = \sum_{\nu=1}^{N_s} {A}^{\nu} B_{\nu}
   = \sum_{\nu=1}^{N_s} {A}_{\nu} B^{\nu}
     \, ,
\ee
which is the discretized version of Eq. (\ref{thisandthat}).

In order to have a discrete Moyal product, we seek to reproduce
the multiplication of operators on the level of symbols. Using the
definition of the symbols, it is straightforward to see that
\be \label{eqProduitMoyal}
(\widehat{A}\widehat{B})_\lambda
      = A_\lambda \ast B_\lambda
      = \frac{1}{(2s+1)^2}
        \sum_{\mu, \nu=1}^{N_s} L^{\mu \nu}_{\lambda} A_\mu B_\nu \, ,
\ee
with the trilinear kernel
\be \label{disctri}
L^{\mu \nu}_\lambda
    = \tr \left[\widehat{\Delta}^\mu \widehat{\Delta}^\nu
    \widehat{\Delta}_\lambda \right]\, ,
\ee
in close analogy to Eq. (\ref{eqPMC}).

\subsection*{Discrete $P$- and $Q$-symbols}
\label{anBaseCoherent}

A particularly interesting set of symbols emerges if, for a given
allowed constellation, only one of the coefficients in the
expansion (\ref{eqNoyDiscECGD}) is different from zero, $\Delta_m
= \delta_{ms}$, say. Then, the kernel consists of $N_s$ operators
projecting on coherent states,
\be
  \widehat{Q}_{\nu} = \ket{{\bf n}_{\nu}} \bra{{\bf n}_{\nu}} \, .
  \label{eqDefQ}
\ee
This is obviously the non-redundant counterpart of Eq.
(\ref{contcohpro}) implying that a self-adjoint operator $\widehat
A$ is determined by a symbol which consists of $N_s$ pure-state
expectation values, the discrete $Q$-symbol,
\be \label{eqDefSymbole}
  A_\nu = \tr [ \widehat{A} \widehat{Q}_\nu ]
   = \bra{{\bf n}_\nu} \widehat{A} \ket{{\bf n}_\nu} \, .
\ee
Let us point out that the introduction of discrete symbols has
actually been triggered by the search for a simple method to
reconstruct the density matrix of a spin through expectation
values \cite{amiet+98}. In fact, this problem is solved by Eq.
(\ref{eqDefSymbole}) in the most economic way. If $\widehat A$ is
chosen to be the density matrix $ \hat\rho $ of a spin $ s $, then
the $ \nu $-th component of the $ Q $-symbol equals the
probability of measuring the eigenvalue $ s $ in the direction  $
{\bf n}_\nu $,
\be
     p_s({\bf n}_{\nu}) = \bra{{\bf n}_{\nu}}
               \hat{\rho} \ket{{\bf n}_{\nu}} \, .
\ee
Knowledge of the $N_s$ measurable probabilities $p_s({\bf
n}_{\nu})$ thus amounts to knowing the density matrix
$\hat{\rho}$.

If the $Q$-symbol (\ref{eqDefSymbole}) determines an operator
$\widehat A$,  the values of the continuous Q-symbol of $\widehat
A$ at points different from those of the constellation must be
 functions of the numbers $ (A_1,\ldots, A_{N_s}) $. For a coherent state
$\ket{{\bf n}_0} \neq \ket{{\bf n}_\nu}$, {\em not} a member of the
constellation, this dependence reads explicitly
\be
\bra{{\bf n}_0} \widehat{A} \ket{{\bf n}_0}
  = \frac{1}{2s+1} \sum_{\nu=1}^{N_s} A^\nu
        | \bra{{\bf n}_0} {\bf n}_\nu \rangle |^2 \, .
\ee
Here the $P$-symbol $A^{dual}$ of $\widehat A$ is required,
calculated from its $Q$-symbol by means of (\ref{discrPgen}) once
the matrix
\be \label{eqpartG}
{\sf G}
    \equiv {\sf G}_{\nu \nu^\prime}
     = \tr \left[ \widehat Q_\nu \widehat
         Q_{\nu^\prime} \right]
     =| \bra{{\bf n}_\nu} {\bf n}_{\nu^\prime} \rangle |^2 \, ,
\ee
has been inverted. Furthermore, knowledge of ${\sf G}^{-1}$
provides immediately the dual kernel  ${\widehat Q}^\nu$ via
(\ref{forth}) but no explicit general expression such as
(\ref{PQkernels}) is known.

It will be shown now how to directly determine the matrix elements
of the dual kernel without using the inverse of ${\sf G}$. The
orthogonality of the kernel and its dual, Eq. (\ref{eqDefOrthoBD})
can be written as
\be \label{findortho}
   \delta_\nu^{\nu^\prime}
      =  \frac{1}{2s+1} \tr
         \left[ \widehat{Q}_{\nu} \widehat{Q}^{\nu^\prime} \right]
      =  \frac{1}{2s+1} \sum_{m,m^\prime=-s}^{s}
                 \bra{m^\prime} \widehat{Q}_\nu \ket{m}
                 \bra{m} \widehat{Q}^{\nu^\prime} \ket{m^\prime} \, ,
\ee
using the completeness relation for the $z$ eigenstates $\ket{m,
{\bf n}_z }$. Introduce an $(N_s \times N_s)$ matrix ${\sf Q}$
with elements ${\sf Q}_{\nu, mm^\prime} = \bra{m} \widehat{Q}_\nu
\ket{m^\prime}$, where the index $(m,m^\prime)$ of the columns
runs through $N_s$ values according to
\be
\left\{(2s,2s), (2s,2s-1),\ldots,(2s,0), (2s-1,2s),\ldots, (0,0)
        \right\} \, .
\ee
As is obvious from (\ref{findortho}), the matrix elements of the
dual kernel, ${\sf Q}^{\nu}_{mm^\prime} = \bra{m} \widehat{Q}^\nu
\ket{m^\prime}$ can be read off once the inverse of the matrix
${\sf Q}$ has been found. The expansion coefficients of a coherent
state $\ket{\bf n}$ in the $z$ basis $\ket{m ,{ \bf n}_z}$ are
given by
\begin{equation}
  \braket{m, {\bf n}_z}{ {\bf n}}
  = \frac{1}{(1+|z|^2)^s} \left(\begin{array}{c} 2s \\  s-m \end{array}
   \right)^{1/2} {z}^{s-m} \, ,
\label{eqProdScalC}
\end{equation}
where the complex number $z$ is the stereographic image in the
complex plane of the point ${\bf n}$ on the sphere. Therefore, one
can write $\sf Q$ as a product of three matrices two of which are
diagonal: ${\sf Q} = {\sf D}_1 {\sf N} {\sf D}_2$. The diagonal
matrices
\bea
  {\sf D}_1  & = &
      {\rm diag} \left[ (1+|z_\nu|^2)^{-2s}\right] \, ,    \\
  {\sf D}_2  & = &
      {\rm diag} \left[
            \left( \begin{array}{c} 2s \\ 2s-m \end{array} \right)^{1/2}
            \left( \begin{array}{c} 2s \\
                    2s-m^\prime \end{array} \right)^{1/2}\right] \, ,
\eea
with $\nu=1,\ldots, N_s,$ and $ m,m^\prime  =0,\ldots, 2s$, have
inverses since all diagonal entries are different from zero. The
hard part of the inversion is due to the matrix ${\sf N}$ with
elements
\be \label{eqDefN}
 {\sf N}_{\nu, m m\prime}
           =  z_\nu^{2s-m} \, (z_\nu^*)^{2s-m^\prime} \, ,
\ee
similar to but not identical with to the structure of a
Vandermonde matrix. As discussed in the following chapter, particular constellations give rise to matrices $\sf N$ with inversion formulae
simpler than the general one. Once $\sf N$ has been inverted, the matrix elements of the dual
kernel are given by the rows of the $(N_s \times N_s)$ matrix
\be
  {\sf Q}^{-1} = {\sf D}_2^{-1} \, {\sf N}^{-1} \, {\sf D}_1^{-1} \, .
\ee

For discrete $Q$-symbols, the kernel $L$ in (\ref{disctri}), which
implements the discrete $\ast$ product, has the form:
\be \label{disctriPQ}
L_{\mu \nu \lambda}
    = \tr \left[\widehat{Q}_\mu \widehat{Q}_\nu
    \widehat{Q}_\lambda\right]
    = \braket{{\bf n}_\lambda}{{\bf n}_\mu} \braket{{\bf n}_\mu}
      {{\bf n}_\nu} \braket{{\bf n}_\nu}{{\bf n}_\lambda} \, ,
\ee
which, by using results from \cite{amiet+91}, can be written as
\bea \label{simpleL}
L_{\mu \nu \lambda}
    &=& \frac{1}{4^{2s}}
        (1 + {\bf n}_\mu \cdot {\bf n}_\nu + {\bf n}_\nu \cdot {\bf n}_\lambda
         + {\bf n}_\lambda \cdot {\bf n}_\mu
         + i {\bf n}_\mu \cdot {\bf n}_\nu \wedge {\bf n}_\lambda)^{2s}  \\
    &=& g_0 ( {\bf n}_\mu     \cdot {\bf n}_\nu)^s
        g_0 ( {\bf n}_\nu     \cdot {\bf n}_\lambda)^s
        g_0 ( {\bf n}_\lambda \cdot {\bf n}_\mu)^s
        e^{i s A(\mu \nu \lambda)} \, ,
\eea
where $ g_0({\bf n}_\mu \cdot {\bf n}_\nu)
         = (1+{\bf n}_\mu \cdot {\bf n}_\nu)/2 $, and, defining $
g(\mu \nu \lambda) $ as the term in round brackets of
(\ref{simpleL}), one has
\be
 A(\mu \nu \lambda) = \frac{1}{i} \ln
     \left( \frac{g(\mu \nu \lambda)} {g^*(\mu \nu \lambda)}
     \right) \, .
\ee
Therefore, the phase $A$ has a geometrical interpretation
\cite{amiet+91}: it is the surface of the geodesic triangle given
by the points $ {\bf n}_\mu, {\bf n}_\nu, {\bf n}_\lambda $ .
\section{Constellations}
In this section examples of specific constellations are presented
for which it is possible to prove at least that the Gram matrix
has a determinant different from zero. Furthermore, in some cases
relatively simple expressions for the dual kernel or,
equivalently, for the inverse of the Gram matrix ${\sf G}$ are
obtained. The kernel is supposed throughout to consist of $N_s$
projection operators $\widehat Q_\nu$ on coherent states as given
in (\ref{eqDefQ}). In other words, the focus is on discrete
$Q$-symbols and the $P$-symbols related to them. Note that, once a
constellation has been shown to give rise to a basis in ${\cal
A}_s$, the inversion of its Gram matrix is {\em always} possible
but lengthy (already for a spin $1/2$): express the matrix
elements of ${\sf G}^{-1}$ in terms of the co-factors of ${\sf
G}$.  Four different types of constellations will be discussed
involving randomly chosen points, points on nested cones, on free
cones, and on spirals.
%
\begin{figure}
 \begin{center}
   \epsfig{file=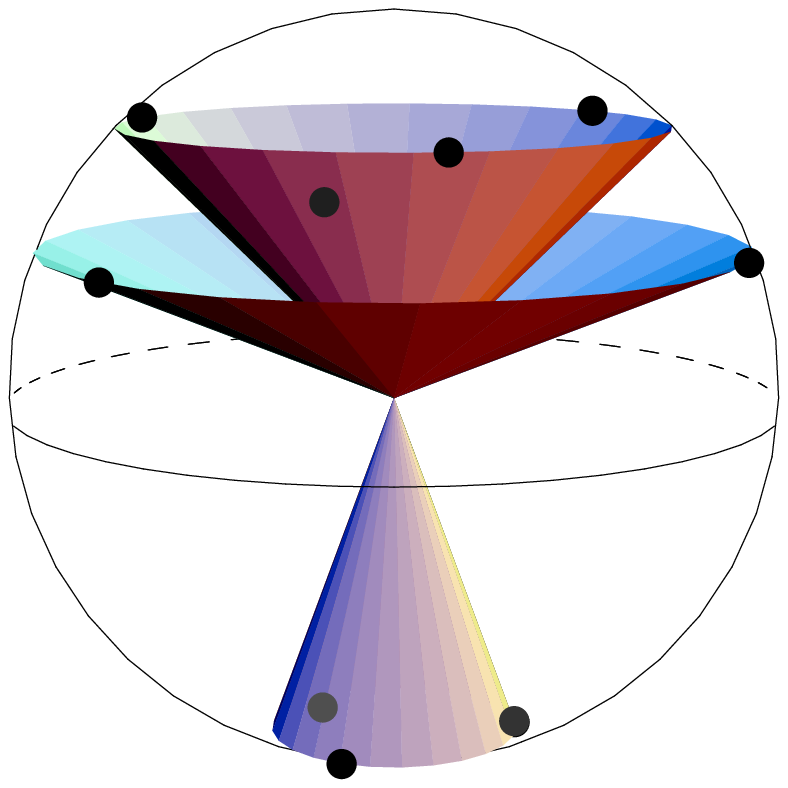,width=4cm}
   \epsfig{file=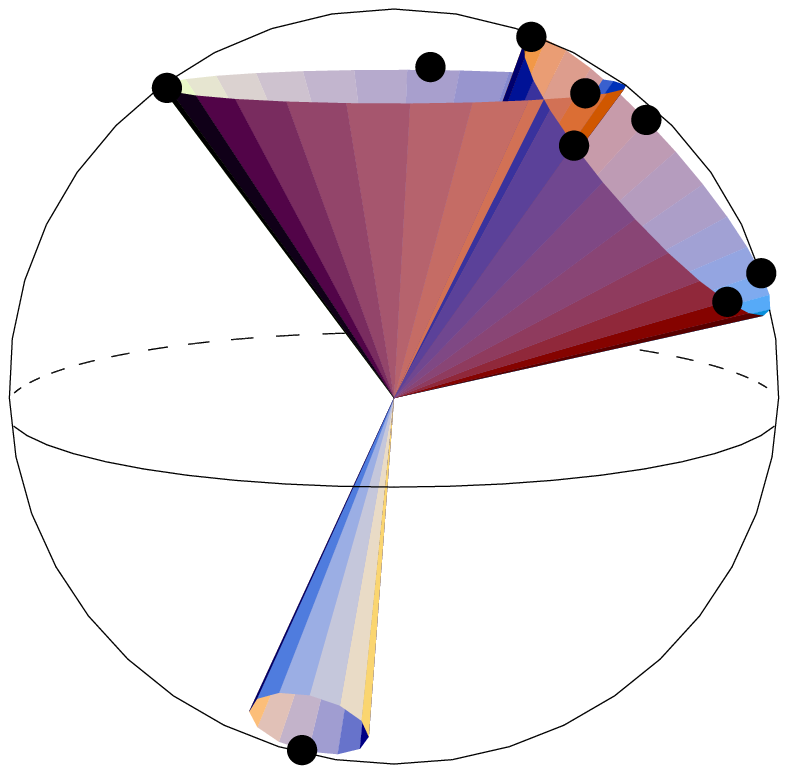,width=4cm}
   \epsfig{file=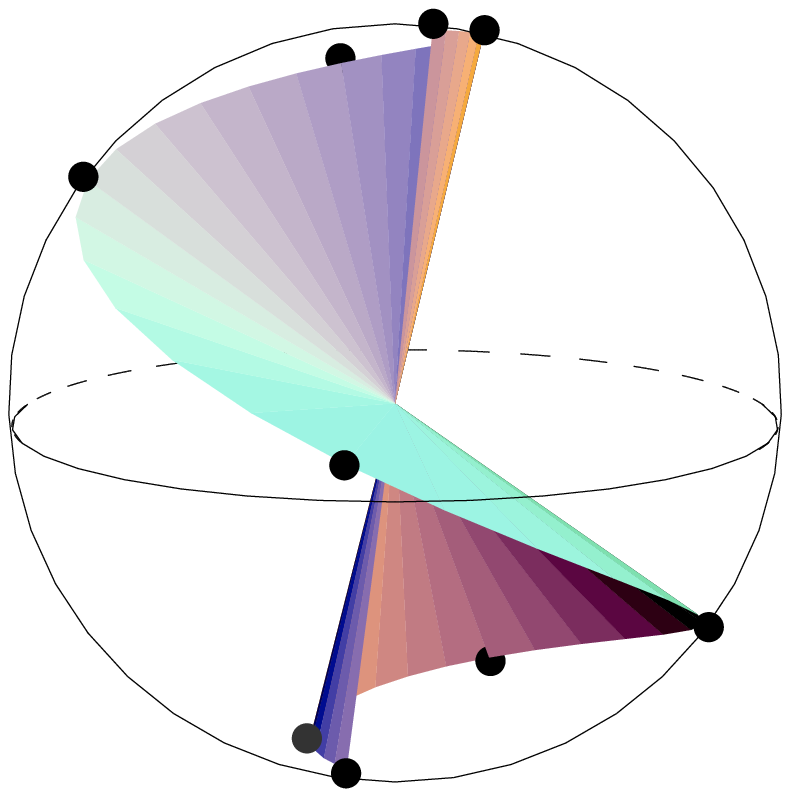,width=4cm}
 \end{center}
 \caption{
   \label{figCons}
   Examples of nested cones, free cones, and a spiral constellation for spin quantum number $s=1$. Each set of nine points defines an allowed constellation.}
\end{figure}
\subsubsection*{Random constellations}
As shown in \cite{amiet+99/3}, almost any distribution of $N_s$
points on the sphere ${\sf S}^2$ gives rise to an allowed
constellation. A random selection of directions leads with
probability one to an invertible Gram matrix. This result shows
that in an infinitesimal neighborhood of any forbidden
constellation one can find an allowed one.
\subsubsection*{Nested cones}
Historically, this family of constellations provided the first
example of allowed constellations  for both integer and
half-integer spins \cite{amiet+99/2}.  For an integer value of
$s$, e.g., consider $(2s+1)$ cones about one axis in space, ${\bf
e}_z$, say, all with different opening angles. Distribute $(2s+1)$
directions over each of these nested cones in such a way that the
ensemble of directions on each cone is invariant under a rotation
about ${\bf e}_z$ by an angle $2\pi/(2s+1)$. For specific opening
angles of the cones, the inversion of the matrix ${\sf N}$ in
(\ref{eqDefN}) reduces after a Fourier transformation to the
inversion of $(2s+1)$ Vandermonde matrices of size $(2s+1) \times
(2s+1)$. For a half-integer spin the same construction is possible
except that the directions on different cones must also lie on
different meridians. There is, in fact, a slight generalization of
this result: the same calculation with $(2s+1)$ {\em arbitrary}
different opening angles leads to $(2s+1)$ {\em generalized}
Vandermonde matrices with nonzero determinant defining thus also a
allowed constellations.

Constellations on nested cones are useful also for numerical
calculations because they allow one to distribute $N_s$ points in
a homogeneous fashion on the surface of the sphere. If two points
of a constellation approach each other, the determinant of the
matrix $ {\sf G} $ typically becomes very large, with a disastrous
effect on numerical precision.
\subsection*{Free cones}
Here is another family of constellation involving $(2s+1)$ cones
with directions located on them. However, now the cones may be
oriented arbitrarily (no nesting), and the number of directions
may vary from cone to cone. For example, the number of points on a cone
can be chosen to equal the multiplicities of the spherical harmonics
$ Y_{lm} $ with $ l =2s$. It is claimed that allowed constellations can be identified by taking into account the following properties (tested numerically for values up to $s =6$):
\begin{enumerate}
  \item   The determinant of ${\sf G}$ is zero if there are more
          than $(4s+1)$ directions on a single cone.
  \item   If there are $(4s+1)$ points on one cone, then another
          cone will contain at most $(4s-1)$ points, allowing for
          no more than $(4s-3)$ directions on the third cone, etc.
  \item   It is necessary to have directions located on at
          least $(2s+1)$ different cones.
\end{enumerate}
For a spin $1/2$ these properties will be shown to hold in the
next section. The first of these observations can be proved for
arbitrary spin $s$ by using a particular decomposition of the
matrix ${\sf G}$,
\be
{\sf G} = {\sf g}^\dagger {\sf g} \, ,
\ee
exploiting the fact that a positive definite matrix can always be
written as the `square' of its `root.' A lengthy calculation
involving properties of rotation matrices, Legendre polynomials
and spherical harmonics leads to a factorization, ${\sf g} = {\sf
d} {\sf y}$, the first matrix being diagonal and having  $(2s+1)$
different entries,      %
\be
{\sf d}_{(l)} = \frac{2 \sqrt{\pi} (2s)!}
    {\sqrt{ (2s+1+l)! (2s-l)!}}  \, ,
    \qquad l = 0, \ldots , 2s \, ,
\ee
each value occurring $(2l+1)$ times. The second matrix has columns
given by the $N_s$ lowest spherical harmonics evaluated at one of
the $N_s$ points of the constellation,
\be \label{ymatrix}
{\sf y} =  \left( \begin{array}{cccc}
       Y_{00} ({\bf n}_1) &  Y_{00}({\bf n}_2)
                          & \dots  & Y_{00} ({\bf n}_{N_s})\\
     Y_{1 -1} ({\bf n}_1) &  Y_{1 -1}({\bf n}_2)
                          & \dots  & Y_{1 -1} ({\bf n}_{N_s})\\
                   \vdots & \vdots
                          &\ddots  & \vdots \\
    Y_{2s 2s} ({\bf n}_1) & Y_{2s 2s}({\bf n}_2)
                          & \dots  & Y_{2s 2s} ({\bf n}_{N_s})\\
           \end{array} \right) \, .
\ee
Consequently, the Gram matrix ${\sf G}$ is invertible if and only
if $\det {\sf y} \neq 0$.  The matrix (\ref{ymatrix}) can accomodate at
most $(4s+1)$ directions on one cone, corresponding
to one value of $\vartheta$ with respect to some fixed axis. The
subsequent multiplicities $(4s-1), (4s-3), \ldots$, are due to applying the same argument to the remaining subspaces with dimensions $2(l-1)+1, 2(l-2)+1,\ldots $

In physical terms, determinant of (\ref{ymatrix}) is easily interpreted as a
Slater determinant of a quantum system: it equals the (totally
anti-symmetric) ground-state wave-function of $N_s$
non-interacting fermions restricted to move on a sphere. The node
lines of this wave function correspond to forbidden constellations
in which the corresponding operator kernel is degenerate, i.e.,
does {\em not} give rise to a basis in ${\cal A}_s$.
\subsection*{Spirals}
A particularly convenient constellation is defined in the
following way: let the $N_s$ directions be defined by $N_s$
complex numbers points $z_\nu$ constructed out of a single point $
z_0 $ (neither of modulus one nor purely real),
\be \label{spiralcoor}
  z_\nu = z_0^{\nu-1}\, , \qquad \nu = 1, \ldots,  N_s \, .
\ee
The points are thus located on a  {\em spiral} in the complex
plane.

The matrix $ {\sf N} $ defined in (\ref{eqDefN}) then turns into
an $(N_s \times N_s)$  Vandermonde matrix, that is,
\be
    {\sf V}_{\nu\mu} = x_\nu^{\mu-1}\, , \qquad \nu,\mu = 1,\ldots, N_s\, .
\ee
Its inverse is known explicitly (\ref{eqInvVDM}), given, for
example, in \cite{verdestar88}, with elements
\be \label{eqInvVDM}
 {\sf V}^{-1}_{\nu\mu}
       = \frac{(-1)^{\nu+1}}{\prod_{\lambda \ne \mu} (x_\lambda-x_\mu)} \,
          S_{N_s-\nu} (\{x_\lambda\}_{\lambda=1}^{N_s} - x_\mu ) \, ,
\ee
 where  $ S_{N_s-\nu}(\{x_\lambda\}_{\lambda=1}^{N_s} - x_\mu) $ is the symmetrical
function constructed out of the $ N-\nu $ numbers $ x_\nu $ with
$\nu \ne \lambda $. One has, for example, $ S_2(x_1, x_2, x_3) =
x_1x_2+x_1x_3+x_2x_3 $.
\section{Discrete Moyal representation for a spin $ 1/2 $}
In this section the discrete Moyal representation will be worked
out in detail for a spin with quantum number $s=1/2$, allowing for
explicit results throughout. For clarity, it is assumed from
the outset that the kernel consists of {\em four projection
operators}
\be
 {\widehat Q}_\nu = \ket{{\bf n}_\nu} \bra{{\bf n}_\nu}\, ,\qquad
 \nu = 1, \ldots, N_s\, .
\ee
It is easy to generalize the results derived below to the case of
four linear combinations of $\ket{\pm{\bf n}_\nu} \bra{\pm{\bf
n}_\nu}$ compatible with Eq. (\ref{eqNoyDiscECG}).

Let us start with the determination of the dual kernel which can
be found by the intermediate step of inverting the $(4 \times 4)$
Gram matrix with elements
\be
{\sf G} = | \bra{{\bf n}_\mu} {\bf n}_\nu \rangle |^2
       =  \frac{1}{2} \left( 1 + {\bf n}_\mu\cdot {\bf n}_\nu
        \right) \, .
\ee
This matrix is easily factorized: ${\sf G} = {\sf g}^\dagger {\sf
g}/2 $, where
\be \label{eqDefA}
  {\sf g} = \left( \begin{array}{c c c c}
                        1 &      1 &      1 &      1 \\
                   n_{1x} & n_{2x} & n_{3x} & n_{4x} \\
                   n_{1y} & n_{2y} & n_{3y} & n_{4y} \\
                   n_{1z} & n_{2z} & n_{3z} & n_{4z}
            \end{array} \right).
\ee
The absolute value of the determinant of ${\sf g}$ is proportional
to the volume of the tetrahedron defined by the four points ${\bf
n}_\nu$ on the surface of the sphere implying $|\det{\sf G}| = 18
V_{tetra}$. Since a `flat' tetrahedron has no volume, the entire
set of forbidden constellations has a simple geometric
description:
\be
  \det {\sf G} = 0
  \Longleftrightarrow
  {\rm the~ four~ points~} {\bf n}_\nu ~
 {\rm are~ located~ on~ a~ circle~ on~} {\cal S}^2.
\ee
Consequently, allowed constellations are characterized by three
vectors on a cone (any three points on a sphere define a circle),
plus any fourth vector not on this cone. This agrees with the
earlier statements about free-cones constellations.

Here is a simple way to invert the matrix ${\sf g}$ and
subsequently ${\sf G}$. Consider a matrix
\be
  {\sf f} = \left( \begin{array}{c c c c}
     1 & {f}^1_{x} & {f}^1_{y} & {f}^1_{z} \\
     1 & {f}^2_{x} & {f}^2_{y} & {f}^2_{z} \\
     1 & {f}^3_{x} & {f}^3_{y} & {f}^3_{z} \\
     1 & {f}^4_{x} & {f}^4_{y} & {f}^4_{z}
            \end{array} \right)\, ,
\ee
defined in terms of four vectors ${\bf f}^\nu = (f_x^\nu, f_y^\nu,
f_z^\nu)$ not required to have length one. The matrix elements of
the of product ${\sf f}$ and ${\sf g}$ are given by
\be \label{almostdiag}
\left({\sf f} {\sf g}\right)^{\mu}_{\nu}
       = 1 +   {\bf f}^\mu \cdot {\bf n}_\nu \, .
\ee
This is a {\em diagonal} matrix if the scalar products $ {\bf
f}^\mu \cdot {\bf n}_\mu$ equal $ -1 $  whenever $\mu\ne\nu $.
Geometrically, such four vectors are constructed easily: the
vector ${\bf f}^1$ points to the unique intersection of the three
planes tangent to the sphere at the points $-{\bf n}_2 $, $ -{\bf
n}_3$ and $-{\bf n}_4$. Analytically, this vector reads
\be \label{eqDefF}
{\bf f}^1 = \frac{{\bf n}_2\wedge {\bf n}_3 + {\bf n}_3
      \wedge{\bf n}_4 + {\bf n}_4 \wedge {\bf n}_2} {({\bf n}_2
         \wedge {\bf n}_3)\cdot {\bf n}_4}\, ,
\ee
and the three remaining vectors follow from cyclic permutation of
the numbers $1$ to $4$. With this choice the inverse of the matrix
${\sf g}$ can be written as
\be
{\sf g}^{-1} = {\sf d}^{-1} {\sf f} \, ,
\ee
where ${\sf d}$ is the diagonal matrix in (\ref{almostdiag}):
${\sf d}_{\nu\nu} = 1+  {\bf f}^\nu \cdot {\bf n}_\nu$.
Consequently, the inverse of the Gram matrix $ {\sf G} $ for a
general allowed constellation is given by
\be
{\sf G}^{-1}
     = 2 \,  {\sf d}^{-1} {\sf f} \,  {\sf f}^\dagger {\sf d}^{-1} \, ,
\ee
having matrix elements
\be \label{eqInverseGspinDemi}
{\sf G}^{-1}_{\mu\nu}
     \equiv {\sf G}^{\mu\nu}
     =  2 \, \frac{1+{\bf f}^\mu \cdot {\bf f}^\nu}
         {(1+{\bf n}_\mu \cdot {\bf f}^\mu)
         (1+{\bf n}_\nu \cdot {\bf f}^\nu)} \, .
\ee
In general, the elements $ {\widehat Q}^\nu $ of the dual kernel
will thus be linear combinations of all four projection operators
$ {\widehat Q}_\nu $.

It is interesting to express the kernel and its dual in terms of
the Pauli matrices $ \bsigma = (\sigma_x, \sigma_y, \sigma_z) $:
\be \label{eqSDQ}
     {\widehat Q}_\nu
      = \frac{1}{2} ( {\sf I}+{\bf n}_\nu \cdot \bsigma)\, ,
      \qquad
     {\widehat Q}^\nu
     = 2 \, \frac{ {\sf I} + {\bf f}^\nu\cdot\bsigma}
                     {1+{\bf f}^\nu\cdot{\bf n}_\nu}\, ,
\ee
allowing one to show easily that they satisfy the required
duality.

For reference, we give the $Q$- and $P$-symbols of the spin
operator
\be
{\bf s}_\nu
     = \frac{1}{2} {\bf n}_\nu \, ,
              \qquad
{\bf s}^\nu
     = \frac{2 \, {\bf f}^\nu}{1+{\bf f}^\nu\cdot{\bf n}_\nu} \, ,
\ee
and of the identity,
\be
I_\nu = 1 \, ,
    \qquad
I^\nu = \frac{4}{1+{\bf f}^\nu \cdot{\bf n}_\nu} \, ,
\ee
and the symbols of arbitrary operators for a spin $1/2$ follow from
linear combinations.
\section{Discussion}
Operator kernels have been used for a systematic study of
phase-space representations of a quantum spin $s$. The kernels
have been derived from appropriate Stratonovich-Weyl postulates
taking slightly different forms for continuous and discrete
representations, respectively. Emphasis is on the {\em discrete}
Moyal formalism which allows one to describe hermitean operators,
including density matrices, by a {\em minimal} number of
probabilities easily measured by a Stern-Gerlach apparatus. As a
useful by-product a natural and most economic method of {\em state
reconstruction} emerges when a quantum spin is described in terms
of discrete symbols. Further, Schr\"odinger's equation for a spin
$s$ turns into a set of coupled linear differential equations for
$(2s+1)^2$ probabilities \cite{weigert99/1}.

In addition, a new form of the kernel defining continuous Wigner
functions for a spin has been obtained (\ref{suchaniceform}): it has
been expressed as
an ensemble of operators obtained from all possible {\em
rotations} of one fixed operator. This is entirely analogous to an
elegant expression of the kernel for particle-Wigner functions as
the ensemble of all possible {\em phase-space translations} of the
parity operator. Therefore, continuous phase-space representations
for both spin and particle systems now are seen to derive from
structurally equivalent operator kernels.

The discrete symbolic calculus is an interesting `hybrid' between
the classical and quantal descriptions of a spin. On the one hand,
this representation is equivalent to standard quantum mechanics of
a spin. On the other, the independent variables carry phase-space
coordinates as labels (\ref{discexp},\ref{discdualexp}). However, only a
finite subset of points in phase space (corresponding to an allowed
constellation) are
involved reflecting thus the discretization characteristic of
quantum mechanics.

The $N_s$ projections operators associated with a constellation of
points define a non-orthogonal basis for hermitean operators
acting on the Hilbert space of the spin. Each projection is a
positive operator, and, altogether, they give rise to a resolution
of unity. One might suspect that they define a positive
operator-valued measure \cite{jauch+67} or POVM, for short.
However, this is {\em not} the case since the closure relation
does not involve just the {\em bare} projections but they are
multiplied with factors some of which necessarily take {\em negative}
values. Such an obstruction through `negative probabilities' is
not surprising; other phase-space representations are based on
quantum mechanical `quasi-probabilities,' known to have this
property, too.

Let us close with a synopsis of the fundamental Moyal-type
representations for particle and spin systems known so far.  \\
\vspace{5mm}
\begin{center}
\begin{tabular}{p{2.5cm}|p{4cm}|p{4cm}|}
 & \ceq{4cm}{self-dual kernel}  & \ceq{4cm}{dual pairs} \\
 \hline
   \ceq{2.5cm}{particle}
 & \ceq{4cm}{Wigner functions \\ {[ unknown ]}}
 & \ceq{4cm}{$P$-, $Q$-symbols \\ {[ unknown ]}} \\
 \hline
   \ceq{2.5cm}{spin}
 & \ceq{4cm}{Stratonovich/Varilly \\ {[ impossible ]}}
 & \ceq{4cm}{Berezin symbols \\ {[ $P^\nu$-, $Q_\nu$-symbols]}} \\
 \hline
\end{tabular}
\end{center}
\vspace{5mm}
The table provides both a summary and points at open
questions.
The individual entries give the names of the familiar {\em
continuous} phase-space representations while the corresponding
quantities for the {\em discrete} formalism are in square
brackets. Future work will focus on developing a discrete
Moyal-type formalism for a quantum particle. To do this, one must
exhibit, for example, a pair of dual kernels one of which would
consist of a countable set of projection operators on coherent
states. This set is required to be a basis in the linear space of
(bounded?) operators on the particle Hilbert-space. It is not
obvious in which way the associate discrete $P$-symbol would
reflect the subtleties of its continuous counterpart which may be
singular. Similarly, the existence of a self-dual discrete kernel
for a quantum particle is an open question.
\subsection*{Acknowledgements}
St. W. acknowledges financial support by the {\em Schweizerische
Nationalfonds}.
%


\end{document}